\documentclass{mem}

\usepackage{natbib}
\usepackage{txfonts}
\usepackage{balance}
\usepackage{graphicx}
\usepackage{flushend}

\begin{document}

\def\teff{$T\rm_{eff }$}
\def\kms{$\mathrm {km s}^{-1}$}

\def\sgon{J0815$+$4729}
\def\sagu{J0023$+$0307}
\def\sbon{J1035$+$0641}
\def\scaf{J1029$+$1729}

\def\skel{J0313$-$6708}
\def\sfre{HE\,1327$-$2326}
\def\schr{HE\,0107$-$5240}


\title{Early phases of the Galaxy from the chemical imprint on the iron-poor stars J0815+4729 and J0023+0307}

\author{
Jonay I. \,Gonz\'alez Hern\'andez\inst{1,2} 
\and David S. Aguado\inst{3,6,1} 
\and Carlos Allende-Prieto\inst{1,2}
\and Adam Burgasser\inst{4}
\and Rafael Rebolo\inst{1,2,5}
}

\institute{
Instituto de Astrof{\'\i}sica de Canarias, E-38200 La Laguna, Tenerife, Spain\\
\email{jonay.gonzalez@iac.es}
\and 
Universidad de La Laguna, Dept. Astrof{\'\i}sica, E-38206 La Laguna, Tenerife, Spain
\and
Dipartimento di Fisica e Astronomia, Universit\'a degli Studi di Firenze, Via G. Sansone 1, I-50019 Seste Fiorentino, Italy
\and
Center for Astrophysics and Space Science, University of California San Diego, La Jolla, CA 92093, USA
\and
Consejo Superior de Investigaciones Cient{\'\i}ficas, E-28006 Madrid, Spain
}

\authorrunning{Jonay I. Gonz\'alez Hern\'andez}

\titlerunning{Chemical abundances of extremely iron-poor stars}

\date{Received: 11 November 2022; Accepted: 20 December 2022}

\abstract{

We have been exploring large spectroscopic databases such as SDSS to search for unique 
stars with extremely low iron content with the goal of extracting detailed information 
from the early phases of the Galaxy. 
We recently identified two extremely iron-poor dwarf stars \sgon~\citep{agu18apjlI} 
and \sagu~\citep{agu18apjlII} from SDSS/BOSS database and confirmed from high-quality 
spectra taken with ISIS and OSIRIS spectrographs at the 4.2m WHT and 10.4m GTC telescopes, 
respectively, located in La Palma (Canary Islands, Spain). 
We have also acquired high-resolution spectroscopy with UVES at 8.2m VLT telescope 
(Paranal, ESO, Chile) and HIRES at the 10m KeckI telescope (Mauna Kea, Hawaii, USA), 
uncovering the unique abundance pattern of these stars,
that reveal e.g. the extreme CNO abundances in \sgon\ with ratios 
[X/Fe]~$> 4$~\citep{gon20apjl}. 
In addition, we are able to detect Li at the level of the lithium plateau in 
\sagu~\citep{agu19apjl}, whereas we are only able to give a Li upper-limit 0.7 dex below 
the lithium plateau in \sgon, thus adding more complexity to the cosmological lithium problem. 
New upcoming surveys such as WEAVE, 4MOST and DESI will likely allow us to discover new
interesting extremely iron-poor stars, that will certainly contribute to our understanding 
of the Early Galaxy, and the properties of the first stars and the first supernovae. 

\keywords{cosmology: observations -- Galaxy: halo -- primordial nucleosynthesis -- stars: 
abundances – stars: Population II -- stars:Population III -- Galaxy:abundances -- Galaxy:formation}
}
\maketitle{}

\section{Introduction}

\begin{figure*}[t!]
\resizebox{12.5cm}{!}{\includegraphics[clip=true]{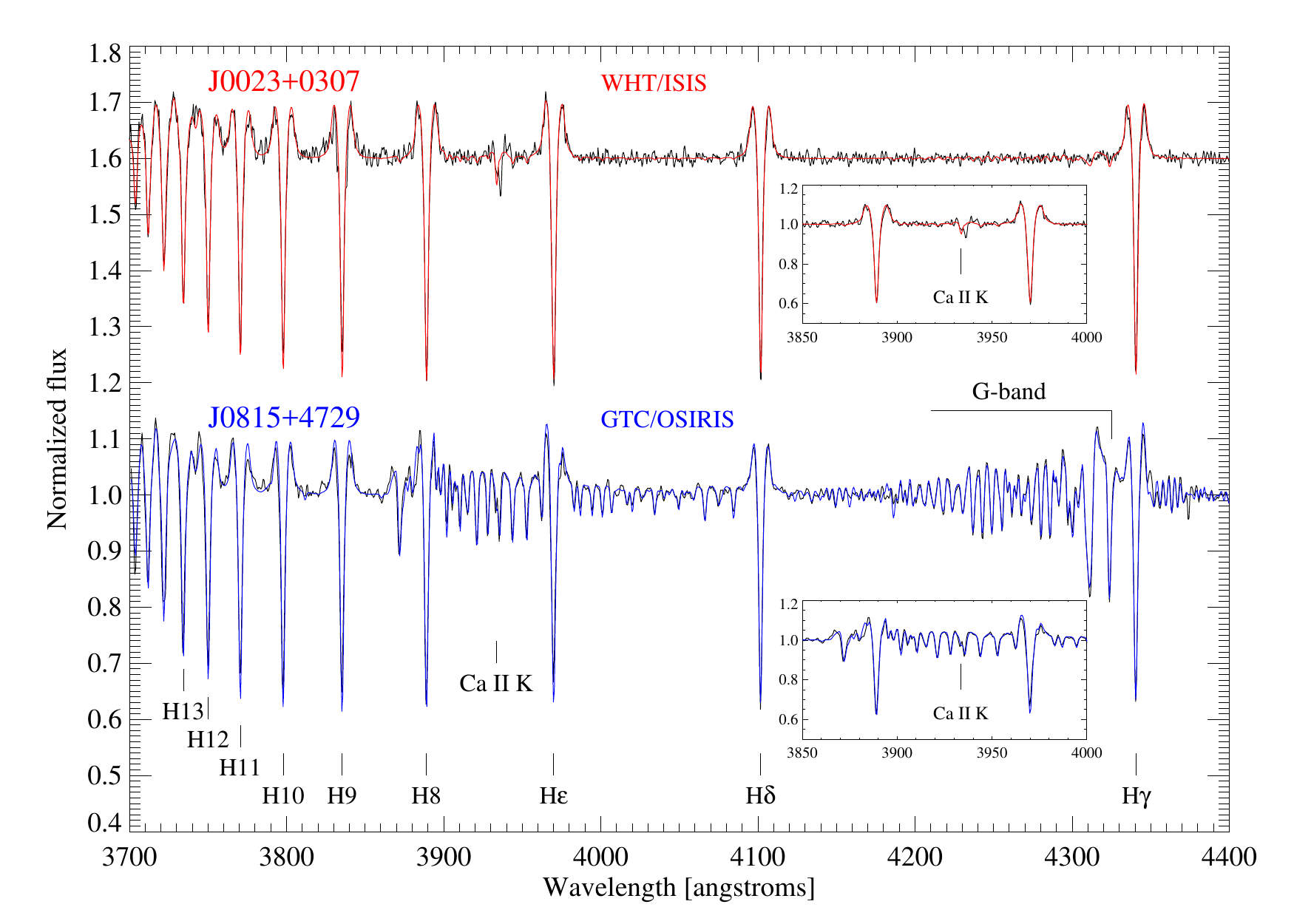}}
\caption{\footnotesize
WHT/ISIS spectrum of J0023+0307 and GTC/OSIRIS spectrum J0815+4729 (black lines) and 
the best fits obtained with FERRE (red and blue lines), normalized using a running-mean
filter. The inner small panels show details of the Ca\,II\,K region for both stars.
}
\label{fig:spec_gtc}
\end{figure*}

Extremely metal-poor stars must have formed from a mixture of material 
from the primordial nucleosynthesis and matter ejected from the first supernovae. 
Those stars are relics of the early epochs of the Milky Way, so their chemical 
composition, especially those still on the main sequence, holds crucial information 
such as the properties of the first stars and the early chemical enrichment of 
the Universe. 

\begin{figure*}
\resizebox{\hsize}{!}{\includegraphics[clip=true]{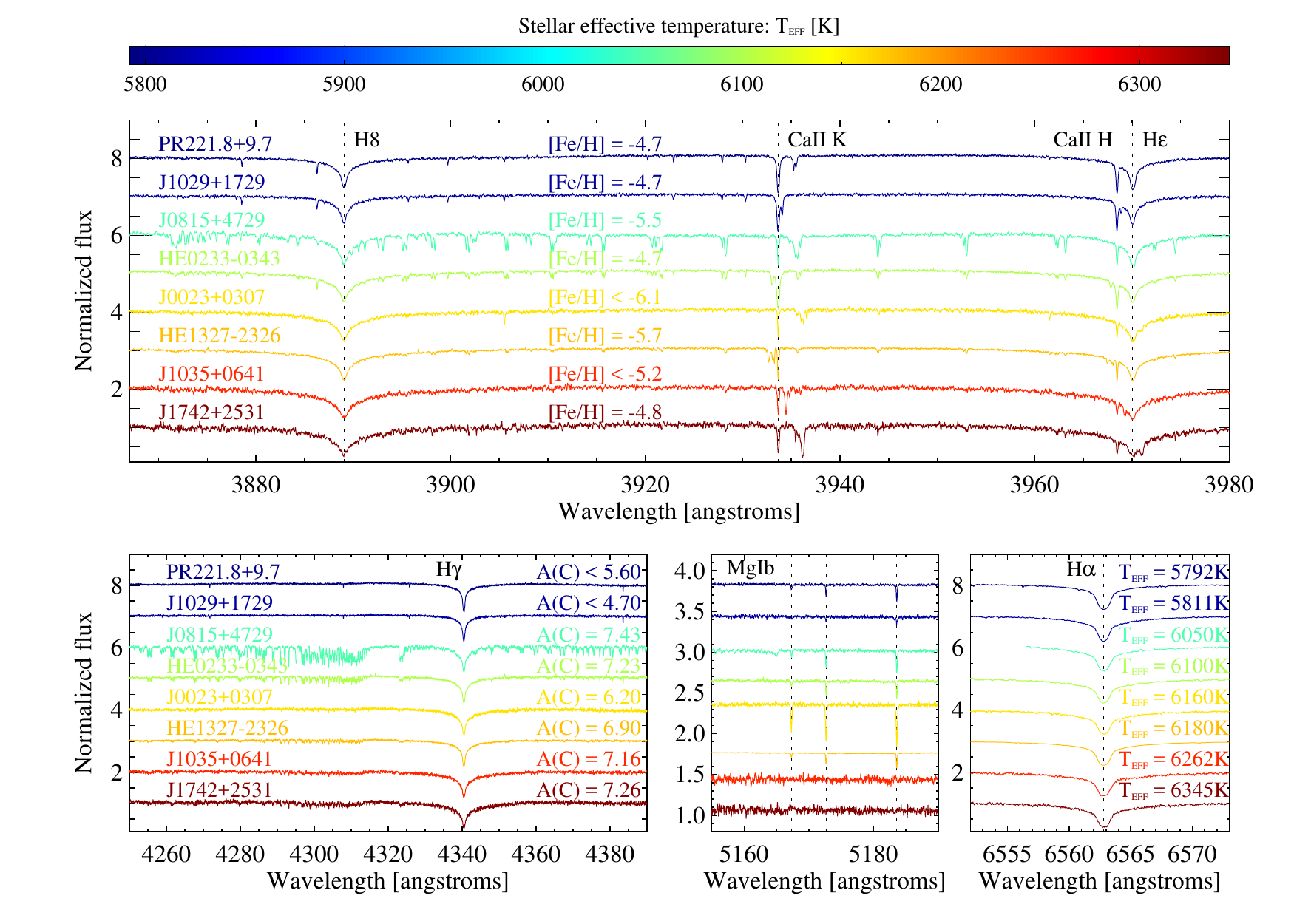}}
\caption{\footnotesize
High-resolution Keck/HIRES spectra of the star J0815+4729 together with the 
VLT/UVES spectra of other extremely iron-poor unevolved stars at [Fe/H]~$< -4.5$. 
The spectra are sorted and colored by stellar effective temperature from top to bottom.
}
\label{fig:spec_all}
\end{figure*}

\begin{figure*}[t!]
\resizebox{7.0cm}{!}{\includegraphics[clip=true]{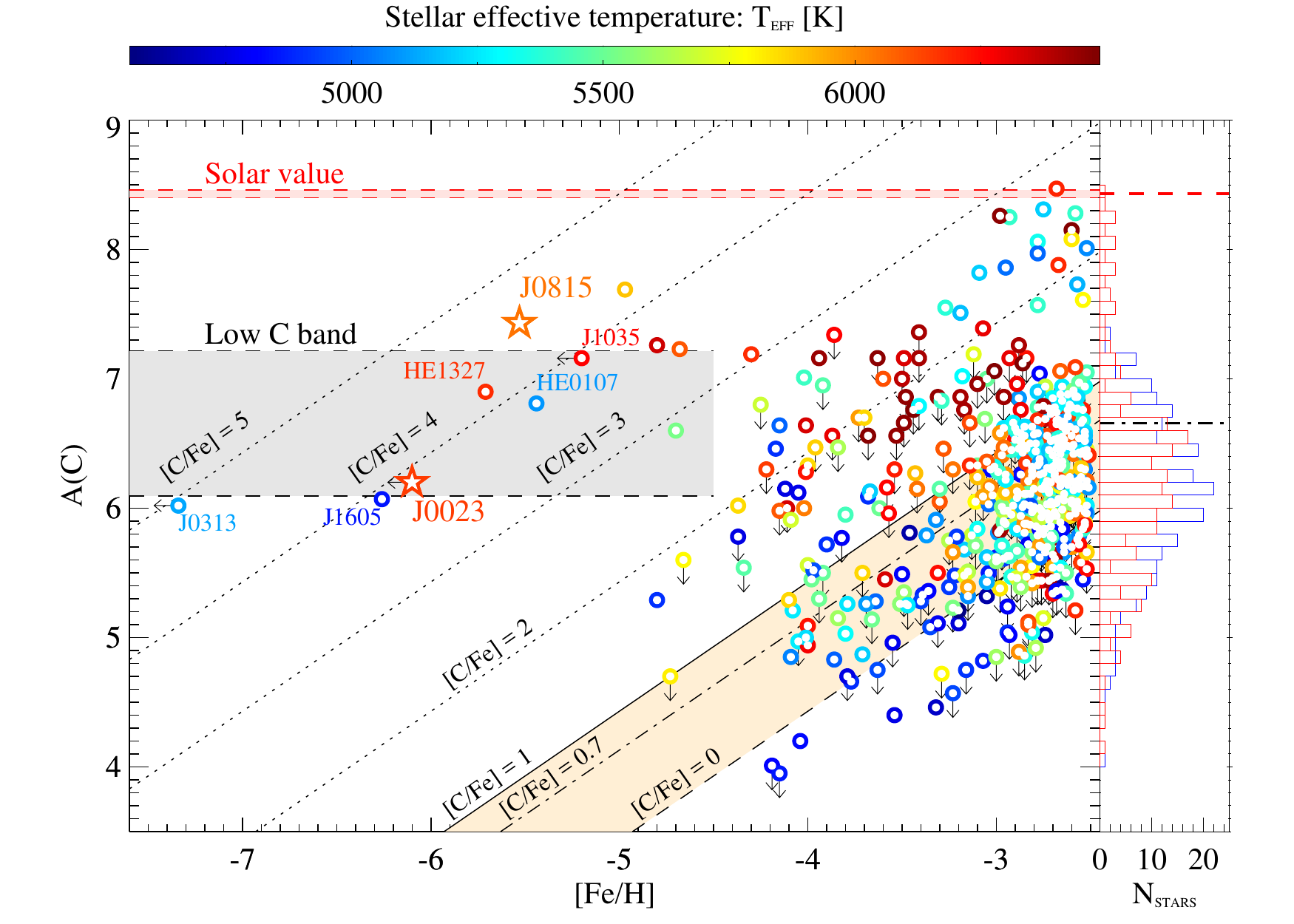}}
\resizebox{7.0cm}{!}{\includegraphics[clip=true]{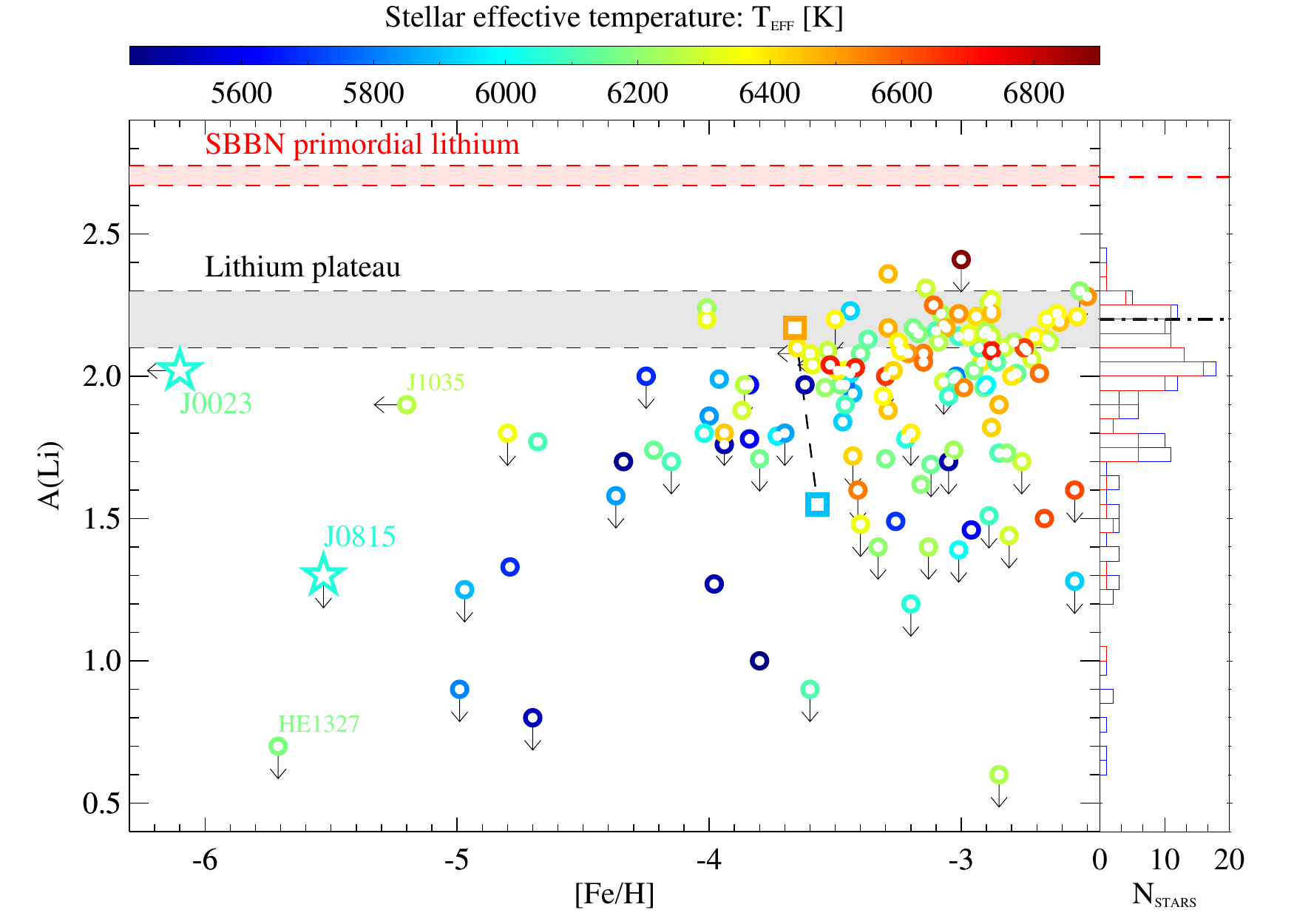}}
\caption{\footnotesize
{\it Left panel}:
Carbon abundance vs. [Fe/H] of the stars \sgon\ and \sagu\ 
(large star symbols) compared with literature measurements for  
metal poor stars with [Fe/H]$< -2.5$. 
{\it Right panel}:
Lithium abundance vs. [Fe/H] of the stars \sgon\
and \sagu\ (large star symbols) compared with literature measurements for unevolved 
metal poor stars with [Fe/H]$< -2.5$ and $\log g > 3.0$. 
The two large squares connected with a black dashed line correspond to the 
metal-poor binary CS 22876–032 AB \citep{gon08aa,gon19aa}. 
In both panels, the literature data include those in 
\citet[][and references therein]{gon20apjl} plus data 
in \citet{yon13apj,han15apj,han16aa,agu19mnras} for C, and those 
new and updated Li abundances in \citet{kie21mnras,lar21mnras,mat21aa}. 
Stars with [Fe/H]~$<-5$ are labelled, including \sagu\ and \sgon.
Symbols are colored by their corresponding effective 
temperature. Downward and left-pointing arrows indicate upper limits. 
}
\label{fig:abun_clife}
\end{figure*}

During the last decades, there has been an enormous observational effort to search
for extremely metal poor stars in large spectroscopic surveys, such as 
Hamburg/ESO~\citep[HE;][]{chr01aa}, or Sloan Digital Sky Survey~\citep[SDSS;][]{yor00aj}, 
or narrow-filter photometric surveys such as Skymapper~\citep{kell07pasa}
or Pristine~\citep{sta17mnras}. However, in the Galaxy with a few hundred thousand 
million stars, 1 over about 800 stars have [Fe/H]~$<-3$ in the solar neighbourhood, 
and we only know 14 stars at metallicity [Fe/H]~$< -4.5$ and only seven at [Fe/H]~$< -5$. 
Almost all stars at [Fe/H]~$< -4.5$ are carbon-enhanced metal-poor (CEMP) stars with 
carbon abundances A(C)~$ > 5$~dex (see Fig.~\ref{fig:abun_clife}), with the clear exception 
of the dwarf star \scaf\ at [Fe/H]~$=-4.7$~\citep{caf11nat}.
At metallicities [Fe/H]~$< -5$, these seven stars appear to be concentrated in the low 
carbon band where all are expected to be CEMP with no enrichment in n-capture 
elements (CEMP-no) with [Ba/Fe]~$< 1$~\citep{bon18aa}. 
Thus, they belong to the CEMP-no class where
their stellar abundances should reflect the pristine material polluted with 
the ejecta of core-collapse supernovae of a few zero metallicity massive stars.

\section{Observations and analysis}

We have extensively explored the SDSS/BOSS~\citep{eis11} 
spectroscopic database and found several tens of extremely metal poor stellar candidates.
Among these we discovered two extremely iron poor stars, \sgon\ and \sagu, that we 
observed using with ISIS and OSIRIS spectrographs at the WHT and GTC telescopes in
the {\it Observatorio del Roque de los Muchachos} (La Palma, Canary Islands, Spain). 
In Fig.~\ref{fig:spec_gtc} we display these very high quality medium-resolution 
WHT/ISIS and GTC/OSIRIS spectra of these two chemically primitive stars that allowed us 
to confirm \sgon\ as an
extreme carbon enchanced star~\citep{agu18apjlI} and \sagu\ as an hyper metal poor with 
apparently no carbon enhancement from the WHT/ISIS spectrum~\citep{agu18apjlII}. 
The GTC/OSIRIS spectrum of \sgon\ shows a forest of CH features together with the series 
of Balmer lines and a tiny Ca\,II K line. The WHT/ISIS of \sagu\ shows also a tiny Ca\,II K
feature but does not show any signature of carbon. We were able to reproduce fairly well 
the observed spectra with synthetic spectral fits using the {\sc FERRE} 
code
\citep[see e.g.][]{agu17aaI}. The global analysis uses FERRE with
a grid of synthetic spectra code ASS$\epsilon$T~\citep{koe08apj} and model 
atmospheres from Kurucz ATLAS 9~\citep{mez12aj}.

\begin{figure}
\resizebox{\hsize}{!}{\includegraphics[clip=true]{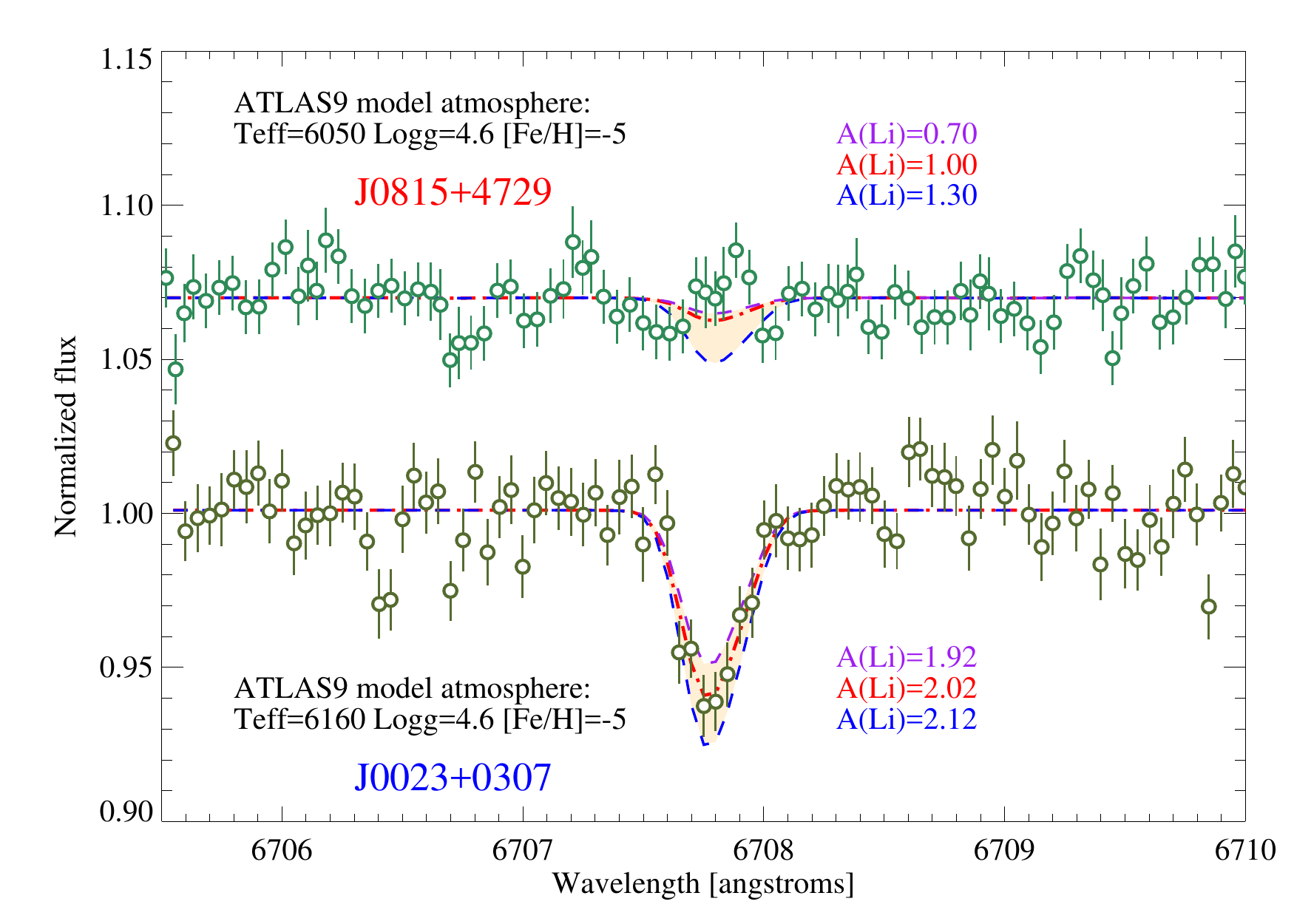}}
\caption{\footnotesize
Lithium features (green dots with error bars) in high-resolution spectra of the star 
\sgon\ (upper Keck/HIRES spectrum) and the star \sagu\ (lower VLT/UVES spectrum) compared 
to SYNPLE synthetic spectra (the best-fit abundance is shown as red dashed-dotted lined).
}
\label{fig:spec_li}
\end{figure}

We observed these two very faint targets using HIRES and UVES at the Keck and 
VLT telescopes, to get high resolution spectra ($R\sim37,500$ for HIRES and 
$R\sim31,000$ for UVES) with the goal of 
extracting the detailed chemical patterns of these two unique stars. A dedicated analysis
of individual spectral features is performed using ATLAS9 model atmospheres and the 1D 
local thermodynamic equilibrium (LTE) using the SYNPLE code for spectral synthesis.  
We also use an automated fitting tool based on the IDL MPFIT routine, with continuum 
location, global shift, abundance, and global FHWM as free parameters~\citep{gon20apjl}.

The individual 1D spectra were corrected for barycentric and radial velocity, normalized, 
merged and binned into a single 1D spectrum of each star. 
In Fig.~\ref{fig:spec_all} we compare these high-quality spectra with those UVES 
spectra of other extremely iron-poor unevolved stars.
Here we see clearly the huge amount of carbon in the spectrum of \sgon\ as compared to 
\sagu\ and other CEMP stars. We also clearly see the Ca\,II\,HK features in all these stars.
We were able to measure a Ca abundance from Ca\,II lines of A(Ca)~$=0.66$ in \sagu\ significantly
lower than the Ca abundance of A(Ca)~$=1.60$ in \sgon, as compared to A(Ca)~$=1.35$ of 
other two extremely iron poor unevolved C-enhanced stars \sbon~\citep{bon15aa} and 
\sfre~\citep{fre08apj}. Given the similarity of stellar parameters of the stars shown in 
Fig.~\ref{fig:spec_all}, the direct comparison of spectra and 1D-LTE element abundances 
seems quite reasonable.
These high-quality spectra allowed us to measure an iron abundance of
[Fe/H]~$=-5.5$ in \sgon\ but only an upper-limit of [Fe/H]~$< -6.1$ (assuming 
[Ca/Fe]~$> 0.4$) in \sagu. Similarly to \skel~\citep{kel14nat} or 
\sbon, no iron lines were detected in the UVES spectrum. We note that the abundance ratio 
[Ca/Fe] in \sgon\ and \sfre\ are 0.75 and 0.71, respectively, which may justify 
the assumption of [Ca/Fe]~$> 0.4$ in \sagu\ and the upper-limit of [Fe/H]~$< -6.1$. 
In Table~\ref{tab:abu} we collect the detailed chemical abundances of 
\sgon~\citep{gon20apjl}, \sagu~\citep{agu19apjl} and \sfre~\citep{fre08apj}.

This UVES spectrum unveiled that the star \sagu\ is indeed
also a CEMP with an abundance from the weak CH G band at the 
$\lambda\lambda 4295-4315$~{\AA} of A(C)~$=6.2$~\citep{agu19apjl}, 
thus providing a high C abundance ratio of [C/Fe]~$>3.9$. 
On the other hand, the HIRES spectrum of \sgon\ is
populated with many C features (see Fig.~\ref{fig:spec_all}), most of them CH lines, 
thus claiming for a significant amount of carbon according to the relatively hot 
effective temperature of the star (only about 100 K cooler than that of \sagu).
This led to the detection of several C molecular features, including CH, CN and C$_2$, 
providing different C abundances. 
We measured inconsistent 1D abundances of A(C)~$=7.4$~dex and~$=8.0$ from CH (G-band) 
and the $\lambda\lambda 5163-5$~{\AA} C$_2$ molecular bands, respectively, as seen 
previously in the iron-poor cool giant \schr~\citep{chr04apj}. We adopted the 
C abundance from CH as the reference C abundance of \sgon, providing a huge C abundance 
ratio of [C/Fe]~$=4.5$. The N and O abundances are also extremely enhanced in \sgon, 
leading to the first detection of the $\lambda\lambda 3881-3$~{\AA} CN molecular band and 
the $\lambda\lambda 7771-5$~{\AA} O~I triplet in an ultra metal-poor unevolved star,
with extreme abundance ratios of [N/Fe]~$=4.4$ and [O/Fe]~$=4.0$~\citep{gon20apjl}.

\begin{table*}
\caption{Element 1D LTE abundances of \sgon, \sagu and \sfre~\label{tab:abu}}
\label{tab:tab1}
\begin{center}
\scriptsize
\begin{tabular}{lrrrrrrrrrr}
\hline\hline
& & \multicolumn{3}{c}{\sgon} & \multicolumn{3}{c}{\sagu} & \multicolumn{3}{c}{\sfre} \\
\cline{3-5}\cline{6-8}\cline{9-11} 
Element & ${\rm A}_\odot({\rm X})^\dagger$ 
& ${\rm A}({\rm X})$ & [X/H]$^{\dagger\dagger}$ & [X/Fe]$^{\dagger\dagger}$ & 
${\rm A}({\rm X})$ & [X/H] & [X/Fe] &  ${\rm A}({\rm X})$ & [X/H] & [X/Fe] \\
\hline
Li & 1.05 &  $<$ 1.30  &       -- &      -- &     2.02 &       -- &       -- & $<$ 0.70 &        -- &       -- \\
C  & 8.43 &      7.43  &  $-$1.00 &    4.49 &     6.20 &  $-$2.23 & $>$ 3.87 &     6.90 &   $-$1.53 &     4.18 \\
N  & 7.83 &      6.75  &  $-$1.08 &    4.41 &       -- &       -- &       -- &     6.79 &   $-$1.04 &     4.67 \\
O  & 8.69 &      7.23  &  $-$1.46 &    4.03 &       -- &       -- &       -- &     6.84 &   $-$1.85 &     3.86 \\
Na & 6.24 &      3.68  &  $-$2.56 &    2.93 &     2.08 &  $-$4.16 & $>$ 1.94 &     2.99 &   $-$3.25 &     2.46 \\
Mg & 7.60 &      3.77  &  $-$3.83 &    1.66 &     4.60 &  $-$3.00 & $>$ 3.10 &     3.54 &   $-$4.06 &     1.65 \\
Al & 6.45 &  $<$ 1.50  & $<-$4.95 & $<$0.54 &     2.35 &  $-$4.10 & $>$ 2.00 &     1.90 &   $-$4.55 &     1.16 \\
Si & 7.51 &  $<$ 3.30  & $<-$4.21 & $<$1.28 &     4.05 &  $-$3.46 & $>$ 2.64 &       -- &        -- &       -- \\
Ca & 6.34 &      1.60  &  $-$4.74 &    0.75 &     0.66 &  $-$5.68 & $>$ 0.42 &     1.34 &   $-$5.00 &     0.71 \\         
Ti & 4.95 &  $<$ 0.70  & $<-$4.25 & $<$1.24 &    -0.42 &  $-$5.37 & $>$ 0.73 &  $-$0.09 &   $-$5.04 &     0.67 \\
Cr & 5.64 &  $<$ 1.50  & $<-$4.14 & $<$1.35 & $<$ 1.54 & $<-$4.10 & $>$ 2.00 & $<$ 0.45 &  $<-$5.19 & $<$ 0.52 \\
Fe & 7.50 &      2.01  &  $-$5.49 &      -- & $<$ 1.40 & $<-$6.10 &      --  &     1.79 &   $-$5.71 &     0.00 \\
Ni & 6.22 &  $<$ 1.90  & $<-$4.32 & $<$1.17 & $<$ 2.93 & $<-$3.29 & $>$ 2.81 &     0.73 &   $-$5.49 &     0.22 \\
Sr & 2.87 &  $<-$1.60  & $<-$4.47 & $<$1.02 & $<-$1.68 & $<-$4.55 & $>$ 1.55 &  $-$1.76 &   $-$4.63 &     1.08 \\
Ba & 2.18 &  $<-$1.40  & $<-$3.58 & $<$1.91 & $<-$1.33 & $<-$3.51 & $>$ 2.59 & $<-$2.14 &   $-$4.32 & $<$ 1.39 \\
\hline
\end{tabular}
\end{center}
Notes: \\
$^\dagger$ Solar abundances from \citet{asp09sun}:
${\rm A}_\odot({\rm X}) = \log \epsilon_\odot({\rm X})=
\log[N_\odot({\rm X})/N_\odot({\rm H})]+12$ \\
$^{\dagger\dagger}$ Element abundances with respect to solar values:  
$[{\rm X}/{\rm H}] = {\rm A}({\rm X})-{\rm A}_\odot({\rm X})$, 
$[{\rm X}/{\rm Fe}] = [{\rm X}/{\rm H}] -[{\rm Fe}/{\rm H}]$\\
\end{table*}

The spectrum of \sagu\ revealed high $\alpha$-element abundance ratios of [Mg/Fe]~$>3.1$ 
and [Si/Fe]~$>2.6$, and also high odd-Z light element abundance ratios of [Na/Fe]~$>1.9$ 
and [Al/Fe]~$>2.0$. The spectrum of \sgon\ shows relatively lower ratios of 
[Mg/Fe]~$=1.7$, [Si/Fe]~$<1.3$, and [Ca/Fe]~$ = 0.75$, but higher ratio of [Na/Fe]~$=2.9$ 
and lower [Al/Fe]$<0.5$. These differences show how unique are the abundance patterns 
of these stars, that are expected to be the result of a mixture of primordial matter 
with the ejecta of a few supernovae of the first massive stars formed in the first 
300 Myr of Universe~\citep{fre15araa}.

\section{Discussion and conclusions}

The detailed abundance patterns from C to Ni permitted a comparison with zero metallicity
SN models, suggesting low-energy SN models with very little mixing of 
21-27~$M_\odot$ Population III progenitors~\citep{heg10apj}. 
The ratios of [Sr/Fe]~$< 1.0$ and [Ba/Fe]~$< 1.9$ in \sgon, do not allow to confirm 
this star as CEMP-no but the carbon abundance is compatible with the upper part of 
the low-carbon band 
(see Fig.~\ref{fig:abun_clife}). The upper-limits in Sr, Ba and Fe do not allow to
extract any conclusion in \sagu\ that also appears to be located in the lower part of 
the low-carbon band. The only few stars known at [Fe/H]~$< -5$ suggest that all
belong to this low-carbon band and may be CEMP-no stars, that are expected
to form at the early phases of the Galaxy and their atmospheric abundances resemble the 
mixture of primordial matter with the ejecta of a few metal-free weak SNe.
There is so far no evidence for RV variations as well as no chemical signature of 
mass transfer from companion AGB stars in \sagu\ and \sgon.
Recently, \citet{agu22aa} have performed a systematic survey of these extremely iron 
poor stars using the ultra-stable high-resolution ESPRESSO 
spectrograph~\citep{pep21aa}. ESPRESSO observations demonstrated the binarity of 
the cool iron-poor giant star \schr\ at [Fe/H]~$= -5.4$ but found a very 
high $^{12}$C/$^{13}$C ratio, thus supporting that this star remains as an unmixed 
CEMP-no with A(C)~$= 6.8$ located in the low-C band.

Finally, these iron-poor stars allow us to look back to the time to the big bang through 
the lithium abundances, in particular, in unevolved stars where Li can still survive 
in their atmospheres during the whole age of the Universe (see Fig.~\ref{fig:spec_all}). 
The star \sagu\ is particular interesting because it shows a significant Li feature at 
6707.8~{\AA}, with a metallicity of [Fe/H]~$< -6.1$, thus 
the only star with a clear Li detection in this metallicity regime (see Fig.~\ref{fig:spec_li}). 
The Li abundance of A(Li)~$=2.0$ in \sagu\ is at the level of the Li plateau 
(see Fig.~\ref{fig:abun_clife}), 
thus extending the upper envelope of the Li abundances in metal poor stars downwards 
by several order of magnitude in metallicity, and keeping unresolved the discrepancy 
between the Li abundances in metal poor stars~\citep[A(Li)~$\sim 2.2$;][]{spi82nat,reb88aa} 
and the primordial Li abundance~\citep[A(Li)~$\sim 2.7$;][]{coc17ijmpe}
predicted from the standard Big Bang Nucleosynthesis at the very beginning of the Universe.

\noindent {\bf{Affiliations}}\par
$^{6}$ INAF-Osservatorio Astrofisico di Arcetri, Largo E. Fermi 5, I-50125 Firenze, Italy\\
 
\begin{acknowledgements}
JIGH, CAP and RR acknowledge financial support from the Spanish Ministry of Science and 
Innovation (MICINN) project PID2020-117493GB-I00. DA acknowledges support from 
the ERC Starting Grant NEFERTITI H2020/808240.

\end{acknowledgements}

\bibliographystyle{aa}

\end{document}